\documentclass[aps,pra,showpacs,twocolumn,superscriptaddress]{revtex4-1}

\usepackage[utf8]{inputenc} 
\usepackage[T1]{fontenc} 
\usepackage{hyperref}
\usepackage{verbatim}
\usepackage[dvips]{graphicx}

\usepackage{gensymb}

\usepackage[version=3]{mhchem}
\usepackage{units}
\usepackage{siunitx}

\usepackage{xspace}

\newcommand{\meta}{\varsigma}
\newcommand{\confLi}{\ce{Li-H}\xspace}
\newcommand{\confH}{\ce{H-Li}\xspace}
\newcommand{\cas}[2]{CAS$^\ast$(#1,#2)\xspace}
\renewcommand{\vec}{\boldsymbol} 

\newcommand{\eq}{Eq.\xspace}
\newcommand{\ie}{\mbox{i.\,e.\ }}

\newcommand{\Ref}{Ref.\xspace}

\newcommand{\DVR}{{DVR}\xspace}
\newcommand{\FEDVR}{{FEDVR}\xspace}

\newcommand{\cre}{\hat a^\dagger}
\newcommand{\ann}{\hat a}

\newcommand{\bfunc}[1]{f_{i_{#1}a_{#1}}^{m_{#1}}}

\newcommand{\matr}[1]{\textbf{#1}}

\newcommand{\matrixe}[3]{\ensuremath{ \left \langle{#1} \left| \vphantom
        {#1 #3} {#2} 
\right| {#3} \right \rangle }}

\newcommand{\smatrixe}[3]{\ensuremath{ \langle{#1} | \vphantom
        {#1 #3} {#2} 
| {#3} \rangle }}

\newcommand{\partd}[2]{\ensuremath{ \frac{\partial {#1}}
{\partial {#2}} }}

\newcommand{\partdd}[2]{\ensuremath{ \frac{\partial^2 {#1}}
{\partial {#2}^2} }}

\newcommand{\ii}{\ensuremath{\mathrm{i}}}

\newcommand{\dd}{\ensuremath{\mathrm{d}}}

\begin{document}

\title{Correlation effects in strong-field ionization of heteronuclear diatomic molecules}

\author{H.~R.~Larsson}
\affiliation{Institut für Theoretische Physik und Astrophysik, Christian-Albrechts-Universität zu Kiel,  D-24098 Kiel, Germany}
\affiliation{Institut für Physikalische Chemie, Christian-Albrechts-Universität zu Kiel, D-24098 Kiel, Germany}
\author{S.~Bauch}
\affiliation{Institut für Theoretische Physik und Astrophysik, Christian-Albrechts-Universität zu Kiel,  D-24098 Kiel, Germany}
\author{L.~K.~S\o{}rensen}
\affiliation{Institut für Theoretische Physik und Astrophysik, Christian-Albrechts-Universität zu Kiel,  D-24098 Kiel, Germany}
\affiliation{Department of Chemistry - {\AA}ngstr{\"o}m Laboratory, Uppsala University, SE-751 20 Uppsala, Sweden}
\author{M.~Bonitz}
\affiliation{Institut für Theoretische Physik und Astrophysik, Christian-Albrechts-Universität zu Kiel,  D-24098 Kiel, Germany}

\date{\today}

\begin{abstract}
 We develop a time-dependent theory to investigate electron dynamics and photoionization processes of diatomic molecules interacting with strong
 laser fields including electron-electron correlation effects.
 We combine the recently formulated time-dependent generalized-active-space configuration interaction theory 
 [D. Hochstuhl and M. Bonitz, Phys. Rev. A {\bf 86}, 053424 (2012); S. Bauch et al., Phys. Rev. A {\bf 90}, 062508 (2014)]
 with a prolate
 spheroidal basis set including localized orbitals and continuum states to describe the bound electrons and the outgoing photoelectron.
 As an example, we study the strong-field ionization of the two-center four-electron lithium hydride molecule
 in different intensity regimes.
 By using single-cycle pulses, two orientations of the asymmetric heteronuclear molecule are investigated:
 \confLi, with  the electrical field pointing from H to Li, and the opposite case of \confH. 
 The preferred orientation for ionization is determined and  we find a transition from \confH, for low intensity, to \confLi, for high intensity.
 The influence of electron correlations is studied at different levels of approximation,
 and we find a significant change in the preferred orientation.  For certain 
 intensity regimes, even an interchange of the preferred configuration is observed, relative to the uncorrelated simulations. Further insight is provided by detailed comparisons of photoelectron angular 
 distributions with and without correlation effects taken into account.

\end{abstract}


\maketitle

\section{Introduction}
The rapid progress in experimentally observing and even controlling electron dynamics in atoms and molecules demands 
powerful theoretical approaches; see, e.g.,~\cite{brabec_intense_2000,kling_attosecond_2008,krausz_2009} for reviews on this subject.
One of the most challenging and, therefore, interesting tasks is the accurate description and understanding of the ultrafast and complex
behavior arising from the electron-electron interaction. It can be expected that mean field, i.e., Hartree-Fock-type approaches are insufficient, and that electronic correlations are important. 
These are especially difficult to treat in time-dependent theories with more than two active electrons, due to the complexity
of the multi-electron wave function and even more so if the continuum is included for photoionization; see~\cite{hochstuhl_2014} for an overview.

Of particular interest in the context of strong field physics are molecular systems due to their much more complex 
dynamics and degrees of freedom owing to their geometrical structure. 
With the development of alignment and even orientation techniques~\cite{stapelfeldt_textitcolloquium_2003,de_field-free_2009,holmegaard_laser-induced_2009,kraus_two-pulse_2014}
measurements in the molecular-fixed frame of reference become accessible, which allows for an investigation beyond orientation-averaged quantities.

One question is the preferred direction of electron emission with respect to the electrical field direction 
of a linearly polarized laser
and the influence of correlation effects in strong-field excitation scenarios of heteronuclear molecules.
In a first approximation, the tunnel ionization maps the highest-occupied molecular orbital (HOMO) to the continuum~\cite{kjeldsen_influence_2005},
and by recollision strong-field ionization was even used to illustrate the HOMO experimentally~\cite{meckel_2008}.

However, experimental evidence using CO molecules~\cite{li_orientation_2011,wu_multiorbital_2012} 
showed that this simplified one-electron picture needs to be adjusted, as effects such as inner-shell polarizations~\cite{zhang_2013},
Stark shifts and orbital distortions~\cite{spiewanowski_alignment_2015} have impact on the ionization dynamics.
The question to what extent electronic correlations are important remains open.
In Ref.~\cite{bauch_2014} this topic has been addressed within a one-dimensional 
model of the four-electron LiH molecule, and a shift of the preferred direction of emission is observed when electronic correlations are included.
The immediate question of whether these effects are present in a full three-dimensional analysis shall be answered by this work and completed
by angle-resolved investigations.

All of these above-discussed issues call for a general, time-dependent theory including external (possibly strong) fields beyond a perturbative approach.
The fundamental equation describing the physics of these quantum systems is the (non-relativistic) time-dependent Schr\"odinger equation (TDSE).
However, its direct numerical solution, even by means of supercomputers, is limited to systems consisting of only one or two electrons,
e.g. helium~\cite{parker_high-energy_2006,taylor_multiphoton_2005,nepstad_numerical_2010,laulan_correlation_2003,feist_nonsequential_2008} 
or molecular hydrogen~\cite{horner_role_2007,colgan_time-dependent_2004}.
Semi-analytical theories, such as  the strong-field approximation
and tunneling theories~\cite{keldysh_ionization_1965, muth-bohm_suppressed_2000,tong_theory_2002, popov_tunnel_2004,tolstikhin_theory_2011,madsen_application_2012}
provide physical insight but often draw on a simplified picture of the electron-electron interactions.

In order to solve the time-dependent Schr\"odinger equation for more than  two  active electrons including the electrons' interactions,
approximate numerical techniques need to be employed. These include the time-dependent configuration interaction singles (TD-CIS) method~\cite{klamroth_laser-driven_2003,krause_time-dependent_2005,gordon_role_2006},
multi-configuration time-dependent Hartree-Fock (MC-TDHF)~\cite{meyer_multi-configurational_1990,nest_multiconfiguration_2005,caillat_correlated_2005,kato_time-dependent_2004,hochstuhl_two-photon_2011}
or its generalizations 
time-dependent restricted or complete active space self-consistent-field (TD-RAS/CAS-SCF)~\cite{miyagi_2013,miyagi_2014,haxton_2015,sato_time-dependent_2013}
and the state-specific-expansion approach~\cite{mercouris_chapter_2010} (see also Ref.~\cite{hochstuhl_2014} for an overview). Further, time-dependent density-functional theory (TD-DFT)~\cite{telnov_2009,chu_comparison_2011} and time-dependent close-coupling
solutions of the TDSE by using pseudo-potentials 
for the description of more than two electrons~\cite{pindzola_time-dependent_2007,pindzola_single_2013} have been applied to photoionization of molecules. 
Especially the MCTDHF family suffers from complicated non-linear numerics, and its applicability to photoionization is not yet fully understood. TD-DFT and the 
pseudo-potential approaches, on the other hand, rely strongly on the chosen functionals or potentials with unknown accuracy and lack tunable parameters to achieve convergence to the fully correlated solution.
One of the most successful methods which bears some similarities to our present approach, is the time-dependent R-matrix method~\cite{van_der_hart_time-dependent_2007,lysaght_2009,van_der_hart_time-dependent_2014}.

The aims of the present work are (i) to provide a fully \emph{ab-initio} time-dependent approach to
electron dynamics in diatomic molecules exposed to strong laser fields including a systematic (i.e.\ controllable) approach to electron correlation without relying on pseudo-potentials
and (ii) to demonstrate the method by shining light onto the question of whether electronic correlation decides from which end an electron leaves a heteronuclear molecule which is exposed to 
a strong electric single-cycle pulse.
Our approach is based on the time-dependent generalized-active-space configuration interaction (TD-GAS-CI) formalism
which we apply within  a prolate spheroidal single-particle basis set in combination with the well-established partition-in-space
concept to tackle the scattering part of the Hamiltonian.

The paper is organized as follows. After a brief introduction into the theory of TD-GAS-CI, we give a detailed overview
on the used basis set and details of our implementation, in Section~\ref{sec:theory}. Technical aspects and the explicit formulas and 
strategies of their efficient numerical handling are presented in the corresponding appendices.
In sections~\ref{sec:results} and \ref{sec:sfi-lih}, we show illustrative numerical examples and demonstrate the abilities of the present approach.
We focus on the LiH molecule in strong single-cycle infrared (IR) pulses and explore the influence of electronic
correlations on the molecular photoelectron angular distributions (PADs) and the preferred direction of electron emission as a function of the
geometrical set-up.
The paper closes with conclusions and a discussion of future applications of the present theory.

\section{Theory}
\label{sec:theory}
Let us consider $N_\text{el}$ electrons moving in the potential of two nuclei at positions $\boldsymbol{R}_A$ and $\boldsymbol{R}_B$ with charge numbers $Z_A$ and $Z_B$.
Throughout, we employ the Born-Oppenheimer approximation~\cite{tannor_book}, 
which decouples the nuclear and electronic degrees of freedom, and use atomic units ($m_e=e=4\pi\epsilon_0=2|E_\text{Ryd}|=a_0=1$).
The (electronic) Hamiltonian is given by
\begin{equation}
 H(t)=\sum_{i=1}^{N_\textup{el}} h_i(t) + \sum_{i<j}^{N_\textup{el}} \frac{1}{|\boldsymbol{r}_i-\boldsymbol{r}_j|} \;,
 \label{eq:Hamiltonian}
\end{equation}
with the one-electron contribution of the $i$-th electron
\begin{equation}
 h_i(t)=-\frac{1}{2} \nabla_i^2+V_i+\boldsymbol{E}(t)\boldsymbol{r}_i \;,
 \label{eq:oneel}
\end{equation}
consisting of the kinetic and potential energies with
\begin{equation}
 V_i=V(\boldsymbol{r}_i)=-\frac{Z_A}{|\boldsymbol{r}_i-\boldsymbol{R}_A|}-\frac{Z_B}{|\boldsymbol{r}_i-\boldsymbol{R}_B|} \;, \label{eq:nuclear_potential_cartesian}
\end{equation}
and the Coulombic electron-electron interaction.
The time-dependent external laser field is denoted by $\boldsymbol{E}(t)$ and is included in dipole approximation using the length gauge via the position operator $\boldsymbol{r}_i$.

\subsection{TD-GAS-CI}
We solve the TDSE for the $N_\textup{el}$ electrons,
\begin{equation}
 i \frac{\partial}{\partial t} |\Psi(t)\rangle = H(t)|\Psi(t)\rangle\;,
 \label{eq:tdse}
\end{equation}
within the TD-GAS-CI framework~\cite{bauch_2014,hochstuhl_2014,hochstuhl_2012,olsen_1988}
with the Hamiltonian~\eqref{eq:Hamiltonian}.
Thereby, we expand the many-particle wave function into a basis of time-independent Slater determinants $|\Phi_I\rangle$,
\begin{equation}
 |\Psi(t)\rangle=\sum_{I \in \mathcal{V}_\text{GAS}} c_I(t) |\Phi_I\rangle \;,
 \label{eq:gas-expansion}
\end{equation}
with time-dependent complex coefficients $c_I(t)$, which results in the matrix representation of the TDSE,
\begin{equation}
 i \frac{\partial}{\partial t} c_I(t)=\sum_{J\in \mathcal{V}_\text{GAS}} H_{IJ}(t) c_J(t) \;.
 \label{eq:tdse-matrix}
\end{equation}
The Slater determinants $|\Phi_I\rangle$ are constructed from single-particle spin orbitals $\chi_i(\boldsymbol{r},\sigma)$ with the spatial coordinate $\boldsymbol{r}$ and
the spin coordinate $\sigma$ and $i=1\dots 2N_b$, where $N_b$ is the dimension of the spatial basis set.
Details of the orbitals are given in Sec.~\ref{ssec:spbasis}.
The matrix elements of the GAS Hamiltonian, $H_{IJ}=\langle \Phi_I| H | \Phi_J \rangle$, can be evaluated either by directly using Slater-Condon rules~\cite{szabo_ostlund_book} or
by efficient techniques from (time-independent) quantum chemistry~\cite{olsen_1988,helgaker_2000}. The most demanding task, besides the time propagation of Eq.~\eqref{eq:tdse-matrix},
remains the evaluation of the one- and two-electron integrals in $H_{IJ}$. Details of our strategy are given in Sec.~\ref{ssec:rot-basis} and App.~\ref{app:integral_trafo}.
The included determinants $\mathcal{V}_\text{GAS}$ in sums~\eqref{eq:gas-expansion} and \eqref{eq:tdse-matrix} are chosen according to the GAS concept described in detail in Ref.~\cite{bauch_2014}.
Thereby, the method ranges from single-active electron (SAE) \cite{schafer_1993,awasthi_2008,hochstuhl_2012} to (exact) full CI.

We propagate Eq.~\eqref{eq:tdse-matrix} using a short-iterative Arnoldi-Lanczos algorithm~\cite{beck_2000,park_1986} which results in the 
repeated application of large-scale matrix-vector multiplications (up to $2\times 10^6$ in one simulation) where the high degree of sparsity of the GAS Hamiltonian
can be efficiently exploited. The algorithm is applied with an adaptive dimension of the Krylov space~\cite{beck_1997,beck_2000} for the propagation of the wave packet with a time-dependent 
Hamiltonian and with an adaptive time-step~\cite{manthe_1991} for the propagation with a time-independent Hamiltonian after excitation with a laser pulse.

\subsection{Single-particle basis}
\label{ssec:spbasis}
For the efficient solution of Eq.~\eqref{eq:tdse-matrix}, a proper single-particle spin-orbital basis $\chi_i(\boldsymbol{r},\sigma)$ 
with an associated spatial orbital  basis $\varphi_i(\boldsymbol{r})$ is required to construct the determinantal basis $|\Phi_I\rangle$.
For quantum-chemistry calculations, typical basis sets are founded on expansions in localized 
functions, such as Gaussian- or Slater-type orbitals (see, e.g.,~\cite{helgaker_2000}). These sets achieve a high precision for bound-state properties but lack an efficient description
of the scattering part of the Hamiltonian.
For atomic targets, i.e., single-center potentials, typically mixed basis sets with radial grids in combination with spherical harmonics are used~\cite{hochstuhl_2012,hochstuhl_2014}. However,
for multi-center geometries, the convergence exhibits unfavorable scaling properties for our case with the required expansion of the angular coordinates~\cite{vanroose_double_2006,tao_2010}.
To overcome this problem for diatomic molecules, we use two-center prolate spheroidal coordinates, in the following.
Alternatively, also for larger molecules with more complicated geometries, 
a combined Gaussian and discrete variable representation (\DVR) can be applied~\cite{rescigno_hybrid_2005,yip_hybrid_2008,yip_hybrid_2014}.

\subsubsection{Prolate-spheroidal coordinates}
\begin{figure}
    \includegraphics[width=8.6cm]{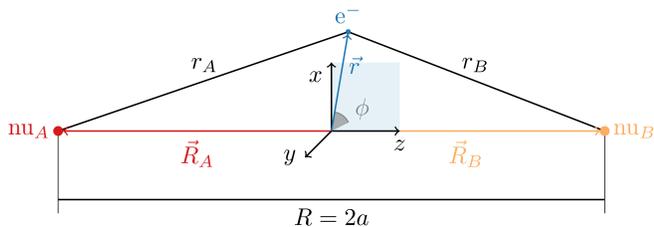}
\caption{(color online). Sketch of the coordinate systems for diatomic molecules~\cite{wahl_1964}. The nuclei are labeled A and B, respectively.}
\label{fig:prol_coords}
\end{figure}
The two-center problem of the diatomic system can be handled efficiently by prolate spheroidal (confocal elliptic) coordinates. 
Therein, we define $a=R/2=|\boldsymbol{R}_A-\boldsymbol{R}_B|/2$ as half the distance between the two centers (nuclei), see Fig.~\ref{fig:prol_coords}:
$r_A$ and $r_B$ are the distances between center $A$ and center $B$  and the electron,
\begin{eqnarray}
  r_A &= &| \vec{r} - \vec{R}_A/2|,\nonumber \\
  r_B &=& | \vec{r} + \vec{R}_B/2|.
\end{eqnarray}
The vectors $\vec r_i$ and $\vec R_i$ ($i\in A,B$) are given in Cartesian coordinates.
The prolate spheroidal coordinates $\xi$ and $\eta$ are then defined as
\begin{align}
 \xi &= \frac{r_A+ r_B}{2a},\;\;\; \xi \in [1,\infty),\\
 \eta &= \frac{r_A-r_B}{2a},\;\;\; \eta \in [-1,1],\\
 \phi &= \arctan\left(\frac{r_y}{r_x}\right),\; \;\; \phi \in [0,2\pi].
 \end{align}
With this definition, center A is located at $z<0$,
 and the binding potential for the electrons, Eq.~\eqref{eq:nuclear_potential_cartesian}, takes the form
 \begin{equation}
   \hat V = - \frac{1}{a (\xi^2-\eta^2)}\left[(Z_A+Z_B) \xi + (Z_B-Z_A)\eta\right].
   \label{eq:pot_diat}
 \end{equation}
Explicit expressions needed for the implementation are comprised in the Appendix.

\subsubsection{Spatial Basis}
For the set-up of the spatial basis, we follow closely Refs.~\cite{tao_2009,
tao_2010,haxton_2011,guan_2011}.
We use a direct product basis where $\xi$ is represented by a Finite-Element \DVR (\FEDVR) basis whose flexibility regarding
the density of grid points avoids complicated coordinate scalings~\cite{balzer_2010,rescigno_2000}.
Coordinate $\eta$ is handled by a usual Gauss-Legendre \DVR \cite{tannor_book}, which is well suited for this problem because spheroidal wave functions are represented by Legendre polynomials.
The spheroidal wave equation is similar to the one-particle Schrödinger equation (see, e.g.,~\cite{meixner_book}).

Although $\hat L^2$ ($\boldsymbol{L}$ is the electronic orbital angular momentum) does not commute with the Hamiltonian, 
$\hat \Lambda$ (component of the electronic orbital angular momentum  along the internuclear axis) does,
and the associated quantum number $m$ is a ``good'' one~\cite{herzberg_book_diatomic}. 
We, therefore, expand the $\phi$-dependent part of $\Psi$ into the eigenvectors of $\hat \Lambda$, which, in our case,
outperformed a Fourier-Grid-Hamiltonian basis in the $\phi$ coordinate~\cite{telnov_2009}:
 \begin{equation}
   \Psi(\xi,\eta,\phi) =\frac1{\sqrt{2\pi}} \sum_{m=-m_\text{max}}^{m_\text{max}} \tilde \Psi^m(\xi,\eta) \exp(\ii m\phi).
  \end{equation}
 
Note that in Refs.~\cite{tao_2009, tao_2010} spherical harmonics $Y_l^m[\arccos(\eta),\phi]$ for the $\eta$ and $\phi$ 
coordinates are used  \footnote{The basis in $\phi$ is then the same basis used in this work.}. This basis 
shows slightly better convergence than a \DVR in $\eta$. However, the resulting electron integrals are less sparse and the
basis is non-orthogonal, which complicates the determinantal basis. 

 To fulfill proper boundary conditions, we use Gauss-Radau quadrature for the first finite element,
which ensures that there is no grid point at the singularity $\xi=1$. 
Gauss-Lobatto quadrature is used for the remaining elements, as usual in \FEDVR~\cite{rescigno_2000}. To avoid singularity of 
the kinetic energy matrix, i.e., to render the matrix invertible for the calculation of the 
two-electron repulsion integrals (see App.~\ref{app:matrix_elements}),
the very last \DVR point of the last element is not included in our grid. 
Thereby an infinite potential barrier at the grid end is created, which
forces the wave function to vanish asymptotically (Dirichlet boundary condition).
If the grid is large enough (i.e., reflections are avoided), effects due to this procedure are negligible.

The one-electron primitive functions used in this work are 
\begin{align}
 f_{ia}^m(\xi,\eta,\phi) &\equiv f_{k}(\xi,\eta,\phi) \nonumber\\
 &= \sqrt{\frac{1}{a^{3}(\xi_i^2-\eta_a^2)}}\theta^m_i(\xi)\theta^m_a(\eta)\frac{\exp(\ii m\phi)}{\sqrt{2\pi}},\label{eq:basis_func_dev}
\end{align}
where a multi-index $k$ has been defined for convenience.
The form of the functions $\theta(x)$ and the matrix elements of the kinetic, potential and interaction energies are given in 
App.~\ref{app:matrix_elements} along with details of their derivation.

\subsubsection{Partially rotated basis}
\label{ssec:rot-basis}
In analogy to Refs.~\cite{hochstuhl_2012,bauch_2014}, we use a partition-in-space concept
to allow for an efficient description of the photoionization process. Similar strategies are also applied in time-dependent $R$-matrix theory, e.g.~\cite{lysaght_2009},
and in Ref.~\cite{yip_hybrid_2010}.
Here the basis set is split at $\xi=\xi_s$ into two parts: an inner region, $\xi <\xi_s$, and an outer part, $\xi\geq\xi_s$.
The splitting point $\xi_s$ is chosen such that it coincides with an element boundary of the FEDVR expansion. This assures 
the continuity of the wave function across the grid and avoids the evaluation of connection conditions, see Ref.~\cite{bauch_2014} for a
detailed investigation in one spatial dimension.

The basis in the inner region, $\xi<\xi_s$, is constructed from Hartree-Fock-like rotated orbitals,
\begin{align}
\phi_i(\vec r) = \sum_{j} C_{ij} f_j(\vec r),
\end{align}
where $\matr C$ is the orbital coefficient matrix.
This rotation is needed because the energy of a truncated CI wave function changes under a unitary transformation of the underlying single-particle basis; 
a good single-particle basis drastically enhances convergence with respect to the size of the truncated CI space, but comes at the cost of
expensive integral transformations which destroy the desired (partial) diagonality of the integral matrices~\cite{helgaker_2000,bauch_2014}. 
Since we are interested in one-electron photoionization with the simultaneous excitation of the ion,
we follow the detailed investigations in~\cite{bauch_2014} and use pseudo-orbitals based on the $N_\text{el}-2$ electron Hartree-Fock problem
for the virtual orbitals in the rotated part of the basis. Here, in contrast to the procedure shown in Ref.~\cite{bauch_2014}, the Hartree-Fock 
problem for the $N_\text{el}-2$ electronic problem is solved with the exchange-potential included, which is appropriate for obtaining localized virtual orbitals. 
The outer part, $\xi\geq\xi_s$, of the basis consists of non-rotated, ``raw'' functions which describe the wave packet in the continuum accurately. One block of the coefficient matrix $\matr C$ is hence diagonal, see Appendix~\ref{subsec:integral_trafo_part_rot_bas}.
An exploitation of the properties of this basis is inevitable for a fast and memory-friendly code [recall that the two-electron
integrals scale, for an arbitrary basis, as $\mathcal{O}(N_b^4)$ whereas for the \DVR basis set they scale as $\mathcal{O}(N_b^2)$]. 
Appendix~\ref{app:integral_trafo} gives details for the efficient transformations of the integrals.

\subsection{Observables}
\label{ssec:observables}
In this part, we discuss the extraction of the relevant observables from the GAS wave function in the mixed-prolate basis set.
More basis-independent details can also be found in Ref.~\cite{bauch_2014}.
\subsubsection{Angular distributions and photoionization yields}
Angular distributions of photoelectrons contain a wealth of information (see, e.g.,~\cite{kumarappan_2008}), 
and, especially in strong fields, dynamical properties of the rescattering process lead to rich structures~\cite{bauch_2008}.
The molecular angular-resolved photoionization yield (photoelectron angular distribution, PAD) is defined as
\begin{align}
 \mathcal{Y}(\theta,t) = \int_{0}^{2\pi}\int_{r_c}^{\infty}\dd \phi \dd r \; r^2 \rho(r,\theta,\phi;t). \label{eq:ionis_yield_theta}
 \end{align}
$\rho$ is the charge density in spherical coordinates ($\theta$ is the azimuthal angle in the $z$-$x$-plane),
\begin{align}
\rho(\vec r;t) &= \sum_{kl}^{N_b} D_{kl}(t) f_k(\vec r)^\ast f_l(\vec r),\label{eq:density_function}
\end{align}
with the spin-summed single-particle density matrix
\begin{align}
 D_{kl}(t) &= \sum_{\sigma} \smatrixe{\Psi(t)}{\cre_{k\sigma}\ann_{l\sigma}}{\Psi(t)}.\label{eq:density_matrix} \;
\end{align}
$\cre_{k\sigma}$ and $\ann_{k\sigma}$ are the creation and annihilation operators, respectively, of a spin orbital with spatial index $k$ 
and spin $\sigma$.
 The critical radius $r_c$ is chosen to be sufficiently large such that only the
``ionized'' part of the charge density is used for integration. This neglects the long-range character of the Coulomb potential
and is strictly  valid only for $r_c \rightarrow \infty$. Therefore, several $r_c$ are used to check convergence. $r_c$ lies usually in the outer region of the partially rotated basis.

The orbitals sampled at grid points in spherical coordinates, $f_k(r,\theta,\phi)$, cf. Eq.~\eqref{eq:density_function},
can be efficiently stored in a sparse vector format by exploiting the locality of the \FEDVR functions. This decreases the memory requirements and the
computation of the charge density~\eqref{eq:density_function} in typical computations by more than three orders of magnitude. 
The photoionization yield at time $t$ can be retrieved from the integrated PAD: 
\begin{align}
 \mathcal P(t) = \int_0^\pi \dd \theta  \mathcal{Y}(\theta,t) \sin \theta.
\end{align}

\subsubsection{Photoelectron energy distributions}

The momentum distribution of the photoelectron is obtained by using the Fourier-transformed basis functions \cite{hochstuhl_2014}. 
Only basis functions outside $r_c$ are used, ignoring the central region. This approach is exact for sufficiently large $r_c$ \cite{madsen_2007}.
The Fourier transform of function $f_k$ is defined as~\cite{arfkenweber}
\begin{align}
 \tilde f_k(\vec p) &= (2\pi)^{-\frac{3}{2}} \int \dd \vec r \exp(-\ii \vec p \cdot \vec r) f_k(\vec r),\label{eq:ft}
\end{align}
where $\vec r$ and $\vec p$ are vectors in prolate spheroidal coordinates and $\vec p \cdot \vec r$ is the inner product in these coordinates.
If the $\phi$-component of $\vec p$ is either $0$ or $\pi$, analytical expressions of \eq~\eqref{eq:ft} can be obtained. However,
the integral kernel is nonanalytic which prohibits the usage of the DVR properties (Gauss quadrature) of the basis functions for the integration.
Therefore, we employ the Fast Fourier Transform of the basis functions in Cartesian coordinates. 
An application of \eq~\eqref{eq:density_function} with the Fourier-transformed basis functions gives then the momentum distribution.

\section{Application to lithium hydride}
\label{sec:results}
Let us consider the diatomic molecule LiH, i.e., $Z_A=3$ and $Z_B=1$.
It is the smallest ($N_{el}=4$) possible he\-te\-ro\-nu\-clear molecule and exhibits a spatial asymmetry with respect 
to its geometrical center.
It is a frequently chosen theoretical model to test correlation methods (e.g., in one spatial dimension in Refs.~\cite{bauch_2014, balzer_2010, balzer_time-dependent_2010, sato_time-dependent_2013,chattopadhyay_electron_2015}).

\begin{figure}
    \includegraphics[width=0.3\textwidth]{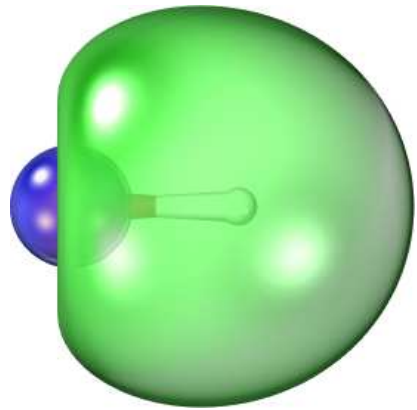}
 \caption{(color online). Isosurfaces of the restricted Hartree-Fock HOMO (orbital to the right, green) and the core orbital (blue, to the left) of the LiH molecule. The Li nucleus is located to the left
 (this corresponds to configuration \confLi in Sec.~\ref{sec:sfi-lih}). 
 The iboview program was used for the generation of the orbitals using an isosurface with a density-threshold of \unit[90.99]{\%}~\cite{iboview}.}
 \label{fig:orbitals}
\end{figure}

We use an internuclear distance of $R=3.015$ ($\unit[1.60]{\textup{\AA}}$), which is the equilibrium geometry at CCSD(T)/cc-pCV5Z level~\footnote{The quantum-chemical computations were performed with the molpro program package~\cite{molpro}.} (without the frozen core approximation) and the experimental value~\cite{herzberg_book_diatomic}. 
The HF electronic structure consists of a valence (HOMO) and a core orbital (1s of lithium), which are given in Fig.~\ref{fig:orbitals}.

The electric field is linearly polarized parallel to the internuclear axis. We consider envelopes of Gaussian shape,
\begin{align}
 E_\text{Gauss}(t) &= E_0 \exp\left[-\frac{(t-t_0)^2}{2\sigma^2}\right] \cos\left[\omega(t-t_0)+ \varphi_{\textup{CEP}}\right], \label{eq:field_gauss}
\end{align}
and of $\sin^2$ shape,
\begin{align}
E_\text{sin$^2$}(t) &= \begin{cases}
E_0 \sin\left(\frac{\omega t}{4}\right)^2\cos(\omega t),& t < \frac{4\pi}{\omega},\\
0,& \text{else},
                       \end{cases}
\label{eq:field_sin2}
\end{align}
with the photon energy $\omega$, the amplitude $E_0$, and the carrier-envelope phase (CEP), $\varphi_{\textup{CEP}}$.

The simulations are carried out within the fixed-nuclei approximation, which is well-justified since the dynamics of the nuclei
are on a much longer timescale than the considered pulse durations.
All data are retrieved by using the length gauge, cf. Eq.~\eqref{eq:oneel}, which is preferable over the velocity gauge in the case of tunnel-ionization dynamics with few-cycle pulses~\cite{madsen_gauge_2002}.

\subsection{GAS partitions}
\label{subsec:results_ci_spaces}
\begin{figure*}
    \includegraphics{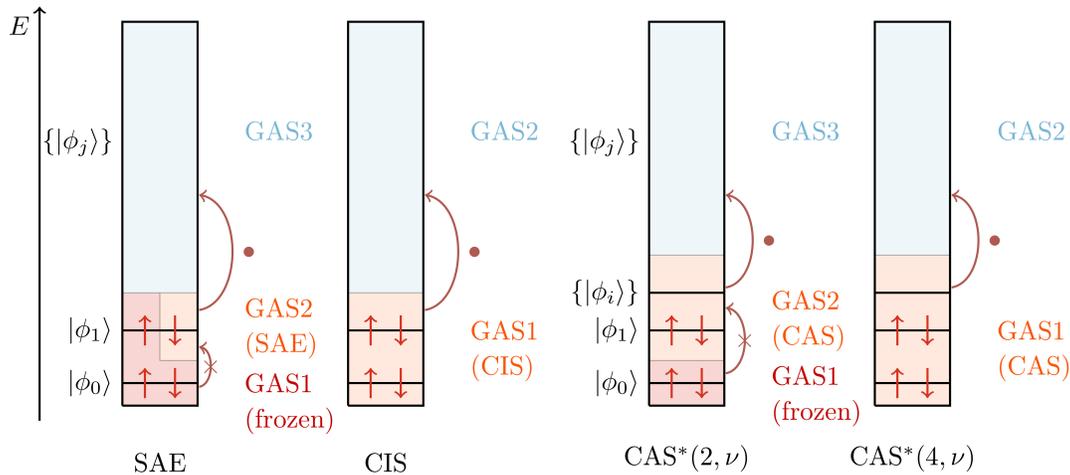}
  \caption{(color online). GAS divisions used in this work. Orbitals (labeled by $\phi_i$) of different spin are assumed to have the same energy.
   The (red) arrows with dots show the allowed excitations. Striked-out arrows mean that no excitations are allowed. The nomenclature is SAE: single-active electron, CIS: configuration interaction singles,
   CAS: complete-active space with $\nu$ spatial orbitals, the star indicates single excitations out of the CAS.}
 \label{fig:gas_cas}
\end{figure*}

The TD-GAS-CI method is well suited for photoionization problems~\cite{hochstuhl_2012,bauch_2014}, as it can be tailored to the problem at hand. 
For constructing the GAS, we assume that multiple ionization is negligible due to the much larger ionization potential of \ce{LiH+}.
We consider three types of GAS partitions which are sketched in Fig.~\ref{fig:gas_cas}:
the single-active electron (SAE) approximation~\cite{schafer_1993,awasthi_2008,hochstuhl_2012},
the CI singles (CIS) approximation and  a complete-active-space (CAS) with single excitations from this subspace 
to the remaining orbitals in the outer region.
Following Ref.~\cite{bauch_2014}, we denote this type of GAS with CAS$^{*}(N_{el}^C,\nu)$, with $N_{el}^C$ describing the number of electrons and $\nu$ the 
number of spatial orbitals in the CAS. The star indicates the single excitations out of the CAS. For the LiH molecule, \cas{2}{$\nu$} describes a CAS with frozen core, and for \cas{4}{$\nu$}, all electrons are active.
As a sidemark, we mention that this type of GAS is equivalent to a multi-reference CIS description.
The size of the active space, i.e., $\nu$ in CAS$^*(\bullet,\nu)$,
remains to be chosen adequately and to be checked carefully for convergence.

For \ce{LiH}, the accurate description of correlation effects requires orbitals with higher quantum numbers $\Lambda$ in the $\phi$ coordinate, i.e., $m_\text{max} > 0$. 
In contrast to time-independent CAS-SCF simulations, where typically a few additional orbitals, such that open shells are filled in the CAS, 
give satisfactory results for the Stark-shifted ground-state energies,
time-dependent calculations are more involved. Here, in order to describe intermediate states accurately, a larger size of the CAS is crucial.
Reasonable CAS configurations contain closed subshells in the number of active orbitals, $\nu$, i.e., all orbitals  with a certain symmetry.
This leads, for example, to a reasonable space of \cas{2}{5} for orbitals with $\Lambda=\Sigma$ and $\Pi$-orbitals for two active electrons [\cas{4}{6} for four active electrons].
Typically, we use CAS spaces with orbitals of up to $\Pi\ (m_{\textup{max}}=1)$, \cas{2}{8} and \cas{2}{12}, 
and $\Delta\ (m_\textup{max}=2)$ symmetry, \cas{2}{10}.

\begin{figure}
    \includegraphics{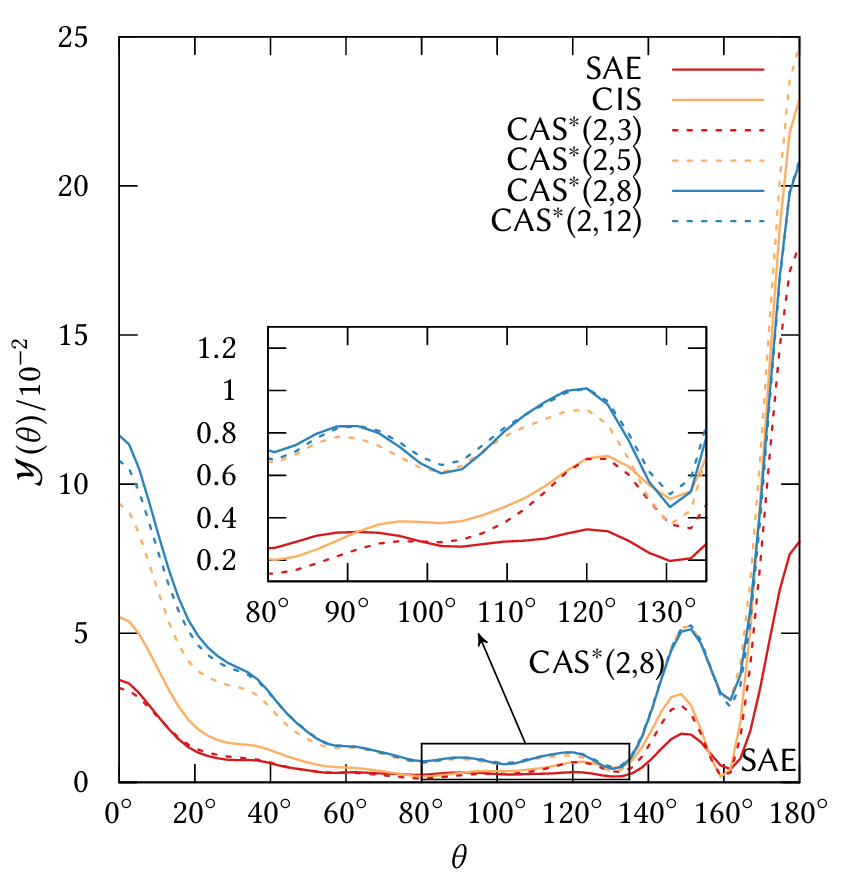}
\caption{(color online). Convergence of the photoelectron angular distribution for different GAS at a field intensity of $\unit[2.2\times 10^{13}]{W\, cm^{-2}}$
with $m_\text{max}=1$. Data is shown for single-cycle excitation of configuration \confH, see Sec.~\ref{sec:sfi-lih} for details.
The PAD for \cas{2}{10} with $m_\text{max}=2$ (\cas{4}{4}) coincides with that of \cas{2}{8} (\cas{2}{3}) and is therefore not shown.}
\label{fig:pad_convergence}
\end{figure}

The convergence of the method is illustrated with an example of the photoelectron angular distributions (PADs) for the
strong-field ionization of LiH using single-cycle pulses in Fig.~\ref{fig:pad_convergence}, see Sec.~\ref{sec:sfi-lih} for the field parameters.
All GAS approximations predict the dominant electron emission in direction of the field polarization, $\theta=0^\circ,180^\circ$.
However, the SAE and CIS approximations drastically underestimate the total yield and fail to predict the correct positions of the side maxima (see inset of Fig.~\ref{fig:pad_convergence}).
The strongest difference is observed in direction of $\theta=0^\circ$. By increasing the active space, a successive convergence is achieved and for a \cas{2}{8} 
only small differences appear in comparison to larger spaces (blue dashed vs. solid blue line). 
Therefore, we typically use the \cas{2}{8} model in the following.
The convergence was checked for different field parameters additionally.

\subsection{One-photon ionization}
\label{sec:one-photon_ionization}
To demonstrate the method, we first consider the case of one-photon absorption in \ce{LiH}. 
The parameters for the electric field, Eq.~\eqref{eq:field_gauss},
are  $\omega = 1.5$ ($\unit[40.8]{eV}$ or $\unit[30]{nm}$), $E_0= 0.005$ ($\unit[0.088\times 10^{13}]{W\, cm^{-2}}$), 
$\sigma = 70$ [$\unit[2.82]{fs}$ full width at half maximum  (FWHM) of  intensity],  
$t_0 = 350$ ($\unit[8.47]{fs}$), and $\varphi_{\textup{CEP}}=0$.
The TDSE in GAS approximation is propagated until $t=900$ ($\unit[22]{fs}$). 
We use up to three functions in $\phi$ ($m_\text{max}=1$), ten in $\eta$, and 1386 in $\xi$. 
The inner region in $\xi$ consists of two elements with 10 and 18 basis functions each and ranges $[1,2]$ and $[2,15)$, respectively.
The non-rotated basis is formed by 80 equidistantly distributed elements with 18 basis functions in each element within $\xi_\text{outer} \in [15,800)$. 

\begin{figure}
    \includegraphics{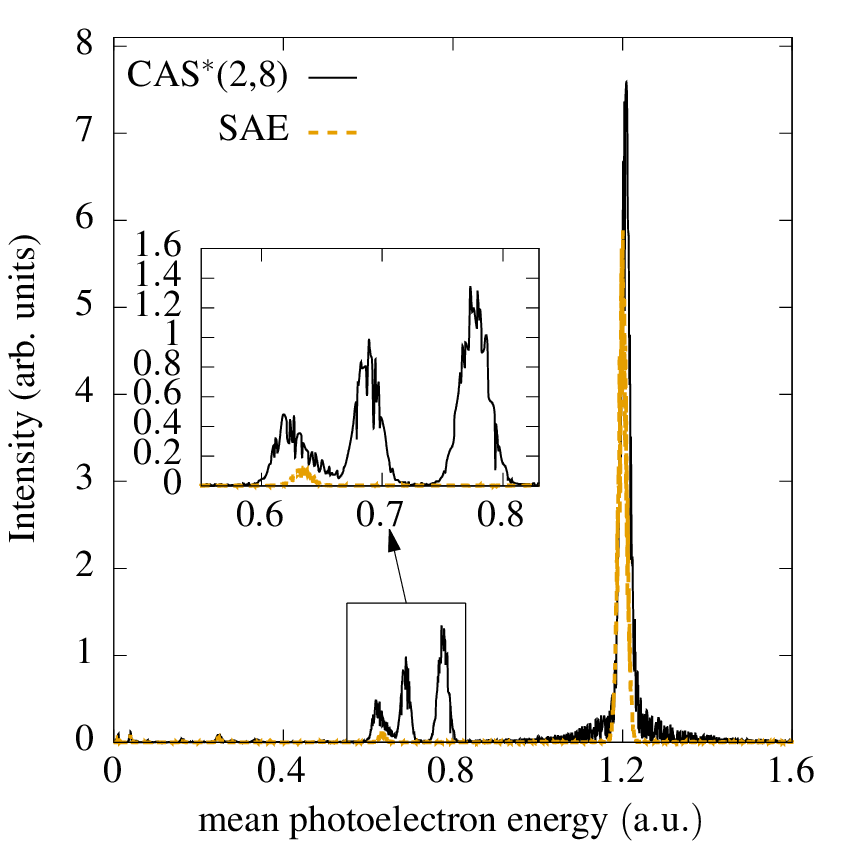}
\caption{(color online). Photoelectron energy spectrum for the ionization of \ce{LiH} 
 with a short ($\unit[2.01]{fs}$) pulse with $\unit[40.8]{eV}$ and an intensity of $\unit[0.088\times 10^{13}]{W\, cm^{-2}}$
 within SAE and \cas{2}{8} (converged) approximation.
 }
\label{fig:distr_resonant}
\end{figure}

The kinetic energy spectrum of the photoelectron is shown in Fig.~\ref{fig:distr_resonant}
for the SAE approximation and the converged \cas{2}{8}.
The spectrum  shows a strong peak at the expected position of 
$E_{\textup{kin}}=\omega-I_p^{(1)}$, where $I_p^{(1)}=|E_v|=0.295$ is the ionization potential according to Koopman's theorem, i.e., the negative
HF energy of the valence orbital, $E_v$.
This peak is rarely shifted by correlations (solid line) but a series of additional peaks appears at lower  kinetic energies (see inset of Fig.~\ref{fig:distr_resonant}, $E<0.8$).
These can be attributed to a correlation-induced sharing of the photon's energy between the photoelectron and a second electron
still bound in the ion (``shake-up'' process). Thereby, the photoelectron energy is reduced and the ion remains in an excited state.
The origin of this process is purely correlation-induced and can neither be described within neither the SAE approach, the CIS approach~\cite{bauch_2014},
nor TD-HF simulations~\cite{bauch_2010}. The population dynamics of these states can be measured, e.g., by strong-field tunneling~\cite{uiberacker_attosecond_2007}.

\begin{figure}
    \includegraphics[width=0.45\textwidth]{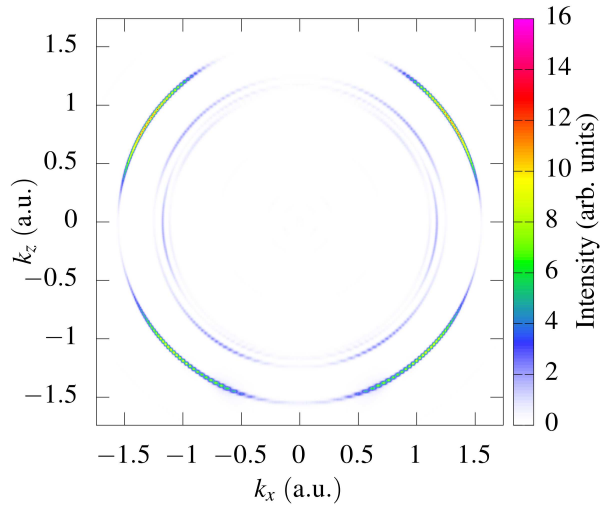}
\caption{(color online). Photoelectron momentum distribution of \ce{LiH} after one-photon excitation for a \cas{2}{8}. 
An integration cutoff of $r_c = \unit[150]{a_0}$ (radial coordinates) was used. See Fig.~\ref{fig:distr_resonant} and text for parameters.}
\label{fig:momentum_distr_resonant}
\end{figure}

The corresponding angle-resolved momentum spectrum depicted in Fig.~\ref{fig:momentum_distr_resonant}, contains additional information.
The photoelectron shows  characteristic  angular distributions for the different peaks with distinct locations of the maxima. The outer circle 
with a radius of about $1.5$ corresponds to the main
photoelectron peak at an energy around $1.2$ in Fig.~\ref{fig:distr_resonant}. The inner (fainter) circles stem from the shake-up state
population and exhibit a significantly different angular dependence than the main peak.
This is caused by the different selection rules for the simultaneous excitation of two electrons and, therefore, the changed angular momentum
of the escaping electron in comparison to the dominant ionization channel with \ce{LiH+} in its ground state.

\section{Strong-field ionization of lithium hydride}
\label{sec:sfi-lih}
We now turn our attention to the case of strong and short pulses, for which a time-dependent theory is indispensable.
Let us consider single-cycle pulses of the form of Eq.~\eqref{eq:field_sin2} with
$\omega = 0.057$ ($\unit[800]{nm}$) which corresponds to a duration of  \unit[5.3]{fs}.

\begin{figure}
    \includegraphics[width=0.48\textwidth]{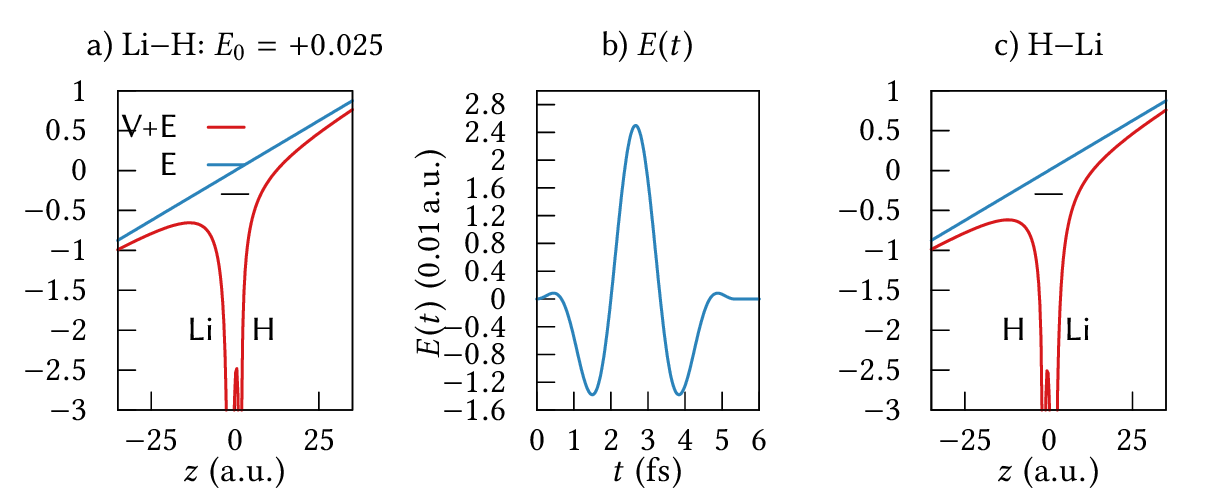}
\caption{(color online). Sketch of the two ionization scenarios of LiH: Figure (a) and (c)
show the potential at the peak of the field at $t_0$ for the configuration \confLi (\confH). The black horizontal line denotes the energy of the active orbital. Panel (b)
shows the time dependence of the single-cycle pulse.}
\label{fig:scenarios}
\end{figure}

The electrical field exhibits a strong CEP dependence with a predominating orientation at the maximum field strength, 
see Fig.~\ref{fig:scenarios} (b). 
This dependence corresponds to two different ionization scenarios, i.e., orientations of the molecule with respect to the field at 
the maximum intensity of the linearly polarized pulse. We will refer to these situations as \confLi, if the field points
from the H to the Li end [panel (a) in Fig.~\ref{fig:scenarios}] and \confH, if the field points from the Li to the H end [panel (c)].

The single-particle basis is similar to Sec.~\ref{sec:one-photon_ionization}, 
but with 16 functions in $\eta$ and 920 functions in $\xi$: 8 and 14 functions in the inner region and 100 elements with 10 functions each in the outer region 
within $\xi_\text{outer} \in [15,1000)$. 
The total number of basis functions in our simulation is $14720$ ($m_\text{max}=0$), $44160$ ($m_\text{max}=1$) and $73600$ ($m_\text{max}=2$).
We further note by comparing the ionization potentials of \ce{LiH} ($I_p^{(1)}=0.295$) and \ce{LiH+} ($I_p^{(2)}=0.825$) that the
Keldysh parameter~\cite{keldysh_ionization_1965} $\gamma=\sqrt{U_p/2 I_p}$ is $1.7$ times larger for the ion. 
Therefore, we conclude that single excitations into the outer region is a valid GAS approximation 
and double ionization is negligible. Convergence aspects of the size of the CAS were discussed in Sec.~\ref{subsec:results_ci_spaces}.

\begin{figure}
    \includegraphics[width=0.48\textwidth]{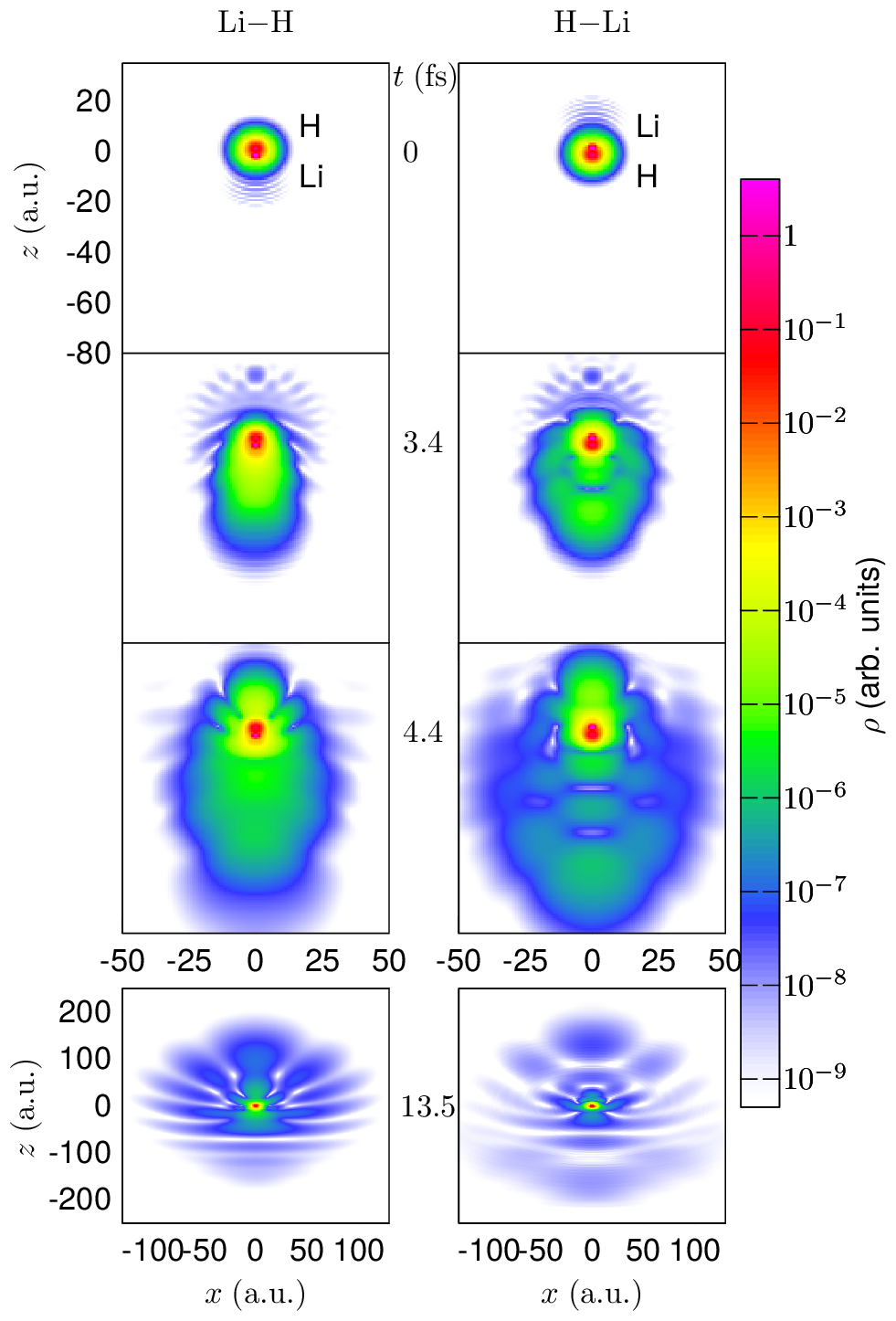}
\caption{(color online). Snapshots of the charge density (logarithmic plot) for the two orientations of the LiH molecule at different times during the 
excitation with the single-cycle pulse using an intensity of $\unit[2.20\times 10^{13}]{W\, cm^{-2}}$
for both scenarios, \confLi (left) and \confH (right), cf.~Fig.~\ref{fig:scenarios}. The classical force $\boldsymbol{F}=-\nabla V$ points downwards.}
\label{fig:charge-density}
\end{figure}
The charge densities at different times during the simulation
for a field strength of $E_1=0.025$ ($\unit[2.20\times 10^{13}]{W\, cm^{-2}}$) and the two scenarios \confLi and \confH are given in Fig.~\ref{fig:charge-density}
for the converged \cas{2}{8} ($m_\text{max}=1$).
The main dynamics happens after about $\unit[3]{fs}$ when the maximum amplitude of the pulse is reached and the tunneling ionization sets in. 
The ejected part of the wave packet exhibits a characteristic angular distribution, visible in the logarithmic density plot. 
After the ionization, the molecular ion remains in Rydberg states and performs
coherent oscillations between the electronic states.
A closer inspection of the released wave packet after the pulse is over ($t=\unit[13.5]{fs}$) reveals a significant difference  for scenarios \confLi and \confH in  both, the angular distribution and the 
absolute yield. The results presented in Fig.~\ref{fig:charge-density} confirm also the observation in Ref.~\cite{bauch_2014} utilizing a one-dimensional model for \ce{LiH} that
ionization for field strength $E_1$ is preferred from the \ce{Li}-end, i.e., using the configuration~\confLi.
This finding will be quantified and discussed in detail in the remaining part of the paper.

\subsection{Orientation dependence of electron emission}
To address the question of whether the ionization yield is larger if the linearly polarized light is pointing
from the \ce{Li}- to the \ce{H}-end (configuration \confH) or \emph{vice versa} (\confLi), the ratio
\begin{align}
 \meta = \lim_{t\rightarrow \infty} \frac{\mathcal P^\text{\confH}(t)}{\mathcal P^\text{\confLi}(t)}
 \label{eq:eta_ionis}
\end{align}
is defined \cite{bauch_2014}. 
$\meta$ is smaller (larger) than one if the electron is ejected mostly in the direction from the  \ce{H}- to the \ce{Li}-end (\ce{Li}- to \ce{H}-end); see also Fig.~\ref{fig:scenarios}.
In Ref.~\cite{zhang_2013} a similar parameter was defined for the CO molecule and a change of the direction for single-active-orbital calculations
in comparison to (uncorrelated) TD-HF calculations including inner orbitals was found.

Let us first consider a field strength of $E_1=0.025$ ($\unit[2.20\times 10^{13}]{W\, cm^{-2}}$).
From the previous discussion, we expect $\meta<1$, since ionization is preferred for configuration  \confLi, cf.~Fig.~\ref{fig:charge-density}.
This can be understood due to the increased total density at the Li nucleus and the direction 
of the classical force $\boldsymbol{F}=-\nabla V$ acting on the electronic density. For the opposite configuration, \confH, 
large portions of the electronic density have to ``pass'' an additional potential well, cf. Fig.~\ref{fig:scenarios}.
This simple picture is confirmed by SAE and CIS calculations in Tab.~\ref{tab:conv_eta} with values of $\meta \lesssim 0.3$.
The total ionization yield is, therefore, about a 
factor of three larger for \confLi than for \confH at field strength $E_1$. 

Additionally,  Tab.~\ref{tab:conv_eta} demonstrates the behavior of $\meta$ with respect to electron correlations by using  different sizes of the CAS.
Similar convergence behavior is also found for other field strengths. 
By successively increasing the CAS, first by including only two active electrons, \cas{2}{$\bullet$},  $\meta$ increases, as it was observed in one-dimensional \ce{LiH}~\cite{bauch_2014}.
Most of the correlation contributions are captured by a \cas{2}{8} with a value of $\meta=0.43$ and only less than 4\% change is found by further increasing the active space.
SAE and the CIS approximations, however, predict values between $0.30$ and $0.32$. Therefore, correlations shift the preferred end of ionization from the Li
to the H end significantly.

We note that too small CAS, e.g.~\cas{2}{3}, give inaccurate results because of a bias due to an improper selection of 
additional important configurations, which is a general pitfall of multi-reference methods~\cite{jensen_book}.
However, to test the frozen-core approximation (correlations arising 
from the two core electrons are not taken into account), we also performed calculations with all four electrons active for a small active space [CIS and \cas{4}{4} in Tab.~\ref{tab:conv_eta}].
By comparing to CIS (frozen core) or \cas{2}{3}, respectively, we find that the observable does not change substantially. 
For larger CAS, a similar behavior is expected, as was demonstrated for one-dimensional systems in Ref.~\cite{bauch_2014}.
Also an increase of $m_{\textup{max}}$ from one to two does not change the result. 
For both, the SAE approximation and CIS, even $m_\text{max}=0$ is sufficient, because the used orbitals in $\phi$ are the corresponding eigenvectors of the one-electron problem.
\begin{table}
 \begin{ruledtabular}
\begin{tabular}{llr}
 Method & $m_\text{max}$ & $\meta$\\
 \hline
 SAE & 0 & 0.30 \\ 
CIS all electrons active & 0 & 0.32\\
CIS frozen core electrons & 0 &  0.31\\
\cas{2}{3}  &1 &  0.31\\
\cas{2}{5}  &1  & 0.35\\
\cas{2}{8}  &1  & 0.43\\
\cas{2}{12} &1  & 0.44 \\ 
\cas{2}{10} &2  & 0.44 \\ 
\cas{4}{4} & 1 &  0.31\\ 
\end{tabular}
 \caption{Ionization asymmetry parameter $\meta$, Eq.~\eqref{eq:eta_ionis}, for the single cycle pulse with field strength $E_1$ ($\unit[2.20\times 10^{13}]{W\, cm^{-2}}$)
 calculated in different GAS approximations with one [SAE/CIS], two [\cas{2}{$\bullet$}] and all four [\cas{4}{$\bullet$}] electrons active.
 $\meta<1$ corresponds to a preferred configuration \confLi for ionization.%
 }
 \label{tab:conv_eta}
 \end{ruledtabular}
\end{table}

\subsection{Intensity dependence}
\begin{figure}
    \includegraphics[width=0.48\textwidth]{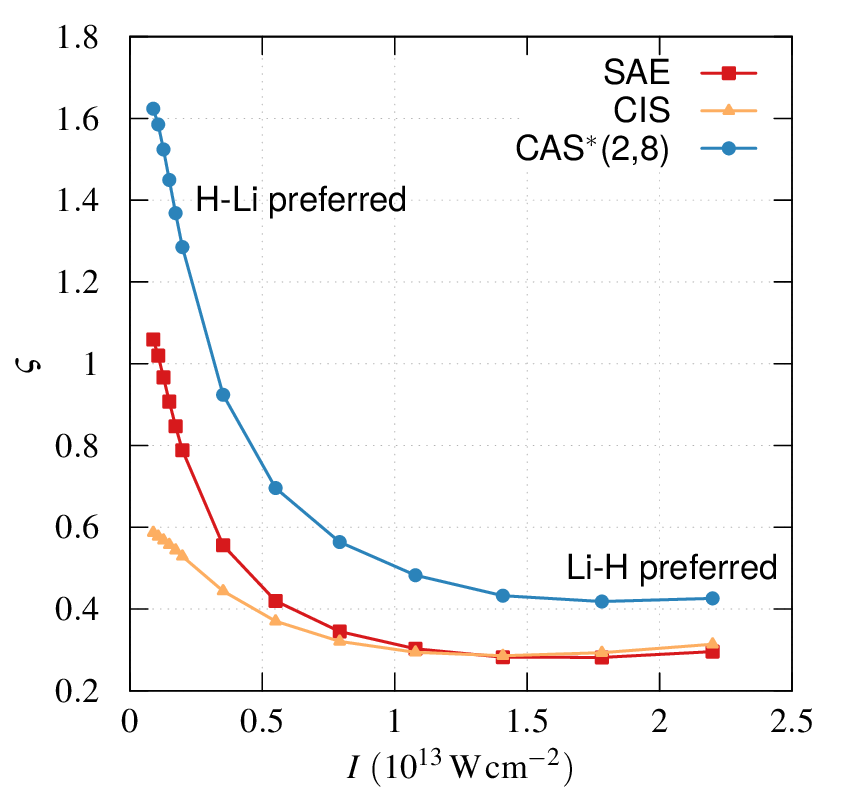}
\caption{(color online). Ratio $\meta$, cf. Eq.~\eqref{eq:eta_ionis}, of the photoionization yields for the two configurations 
[see Fig.~\ref{fig:scenarios}] as a function of the field intensity for the relevant GAS approximations.  $\meta<1$ ($\meta >1$) corresponds to ionization from the \ce{Li}- (\ce{H})-end.
See text for all other parameters.}
\label{fig:meta}
\end{figure}

In the previous paragraph, we found a dominating release for configuration \confLi at a rather high intensity of $\unit[2.20\times 10^{13}]{W\, cm^{-2}}$, which is well above the barrier. This is understandable by the higher electron density at the Li nucleus.
The HOMO, however, is more localized at the H end (green orbital to the right in Fig.~\ref{fig:orbitals}),
in contrast to the full single-particle density. Thus, entering the tunneling regime by reducing the intensity,
the mapping mechanism of the HOMO to the continuum becomes important. Therefore, we expect larger values of $\meta$ for low intensities, 
and thus by increasing the intensity, a decrease of $\meta$; in other words, we expect a shift from the preferred \confH to the \confLi configuration.

This expectation is readily verified by the simple SAE approximation in Fig.~\ref{fig:meta} (red line with squares).
For small intensities, $\meta>1$, exhibiting values of about $1.06$ indicating a slightly  more favorable ionization of the \confH configuration.
For this situation, the CIS approximation (bright yellow line with triangles) fails to describe the change of preferred configuration. A $\meta$ of maximal $0.6$ is predicted by CIS. 
The better qualitative description of the physics by the SAE approximation compared to CIS is probably due to error cancellation.

By increasing the intensity, a monotonic transition to the previously discussed strong-field over-the barrier regime with $\meta<1$ is observed.  
The change of the preferred configuration for ionization from \confH to \confLi ($\meta\approx1$) occurs for the SAE approximation at around $\unit[1\times 10^{12}]{W\,cm^{-2}}$.
For larger intensities, SAE and CIS calculations approximately coincide and the curves approach a ratio of about $0.3$, i.e., a factor of three higher yield for \confLi.
For very high field strengths, the value increases again, which results from the fact that for extremely intense pulses, the asymmetry of the binding
potential is negligible in comparison to the excitation potential and, therefore, a ratio of $\meta = 1$ in the limit $E_0 \rightarrow \infty$ is expectable.

Electronic correlations  [\cas{2}{8}, blue line with circles] change this picture qualitatively, in particular for smaller intensities:
At a specific intensity range from around $\unit[3\times 10^{12}]{W\,cm^{-2}}$ to $\unit[1\times10^{12}]{W\,cm^{-2}}$, the correlation contributions interchange the
dominant direction of emission, $\meta>1$ for \cas{2}{8} and $\meta<1$ for SAE and CIS. At $\unit[1\times10^{12}]{W\,cm^{-2}}$, $\meta$ is much larger ($1.64$) for \cas{2}{8} compared to SAE ($1.06$), indicating that, during the slow tunneling process of the electron from the \ce{H}-end to the continuum, much electron correlation can be built up. 
For large intensities, a similar trend as for the uncorrelated calculation is found, but with
larger absolute values of $\meta$, similar to the case of field strength $E_1$, which is discussed in detail above; see Tab.~\ref{tab:conv_eta}.

\subsection{Photoelectron angular distributions}
\label{ssec:pad}

\begin{figure*}
    \includegraphics[width=0.95\textwidth]{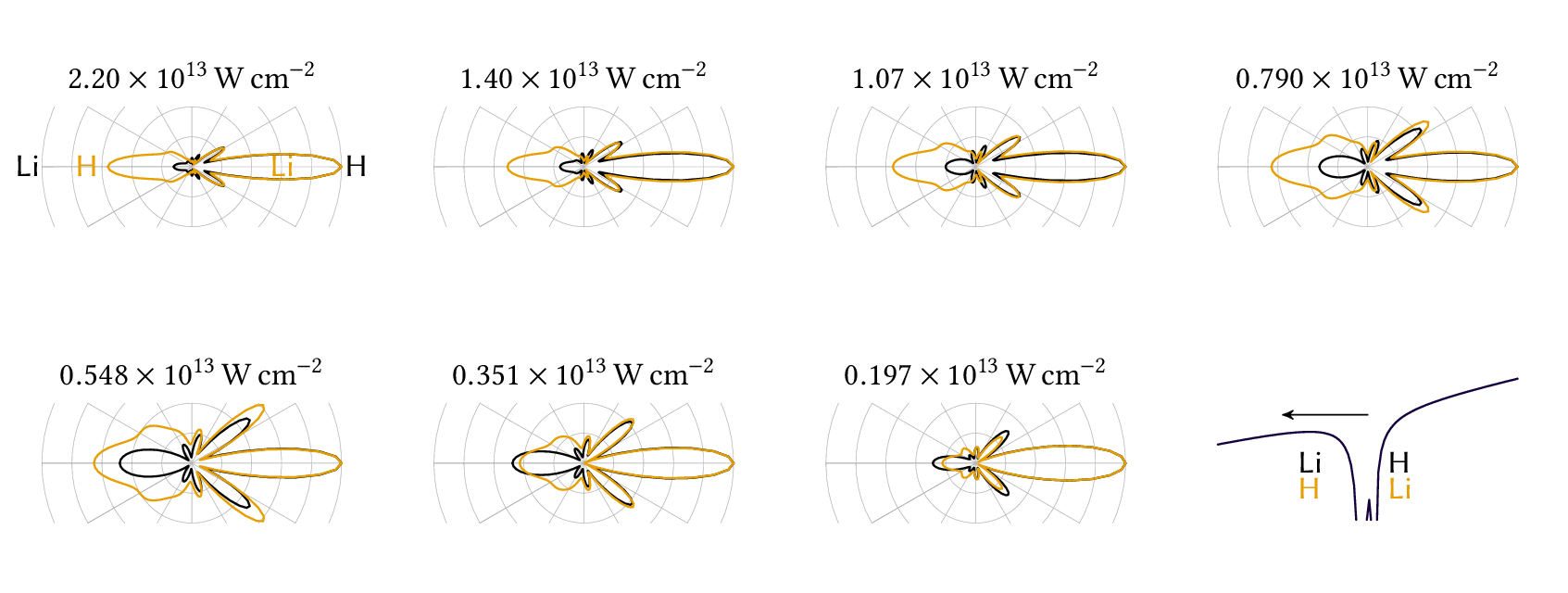}
\caption{(color online). Comparison of the PADs for the two configurations \confLi and \confH at different field intensities for a correlated simulation with \cas{2}{8} using
single-cycle pulses. All distributions have been scaled to fit in the range $[0,1]$.
Note that the schematic (bottom right) gives the potential at the maximum
of the electrical field. The dominant direction of the PAD includes also
the propagation of the electrons in the field after ionization; see the text for a discussion.
}
\label{fig:PADs_field_strengths}
\end{figure*}

We now turn our attention to the fully resolved PADs which contain more information than the integral quantity $\meta$.
The correlated  \cas{2}{8} PADs for the two molecular orientations with respect to the electric field 
at different field strengths are depicted in Fig.~\ref{fig:PADs_field_strengths} in polar plots.
As expected, the dominant ionization occurs along the field polarization axis for all considered intensity regimes.
However, for all field intensities shown,  the most favorable ionization direction is the opposite direction of the field, see sketch at lower right of Fig.~\ref{fig:PADs_field_strengths}.
Further, the ejection direction of electrons differs from the  preferred configuration for ionization measured by the ratio of the integral quantity $\meta$. 
For example, at the highest field strength (top left in Fig.~\ref{fig:PADs_field_strengths}) the PAD
shows a preference of \confH over \confLi, whereas the value of $\meta<1$  prefers $\confLi$ [see Eq.~\eqref{eq:eta_ionis} and Fig.~\ref{fig:meta}]. This can be attributed to the propagation of
the released electrons in the field: whereas $\meta$ depends mainly on the orientation of the molecule with respect to the main peak of the field, the PADs are modified by all cycles of the
pulse and the cycles following the field maximum at which the tunneling-release of the electrons predominantly occur can change the direction of electron emission significantly. 
This is similar to the rescattering mechanism for higher-harmonics
generation and above-threshold ionization. This picture is verified by the time-dependent electron densities in Fig.~\ref{fig:charge-density} 
where after the main peak of the pulse at $t=\unit[3.4]{fs}$, the electron density gets accelerated by the smaller side-extremum in opposite direction (compare with $t=\unit[4.4]{fs}$ in  Fig.~\ref{fig:charge-density}).
A similar effect is, e.g., observed in strong-field ionization of atoms, where rescattering effects can drastically modify the angular distributions of the photoelectrons \cite{bauch_2008}.

With decreasing field strength (top left to bottom right), the shape of the curves along the main maximum become more oblate and the smaller maxima first increase and then decrease again.
For the highest field strengths, the PADs for the configuration \confH (bright yellow curves) show also ionization contributions to the opposite direction pointing to the \ce{H} end, which is not the case for \confLi (black curves). However, with decreasing field strength, the maximum pointing to the other direction is growing for configuration \confLi, and gets even larger than that of \confH. Remarkably, the positions of the secondary
maxima at this intensity are the same regardless of the position of the nuclei, but at intermediate field strengths, an additional maximum for the \confH configuration at the site of the \ce{H} nucleus is visible. This indicates a higher angular momentum for the ejected electron.

\begin{figure*}
    \includegraphics{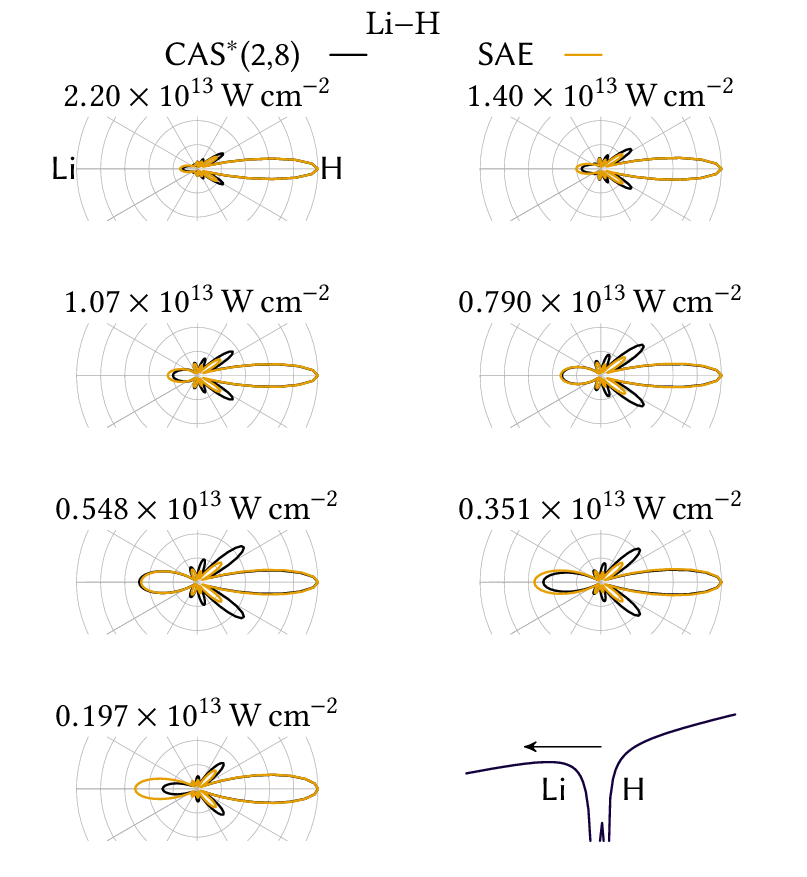}
 \hspace*{1em}
 \includegraphics{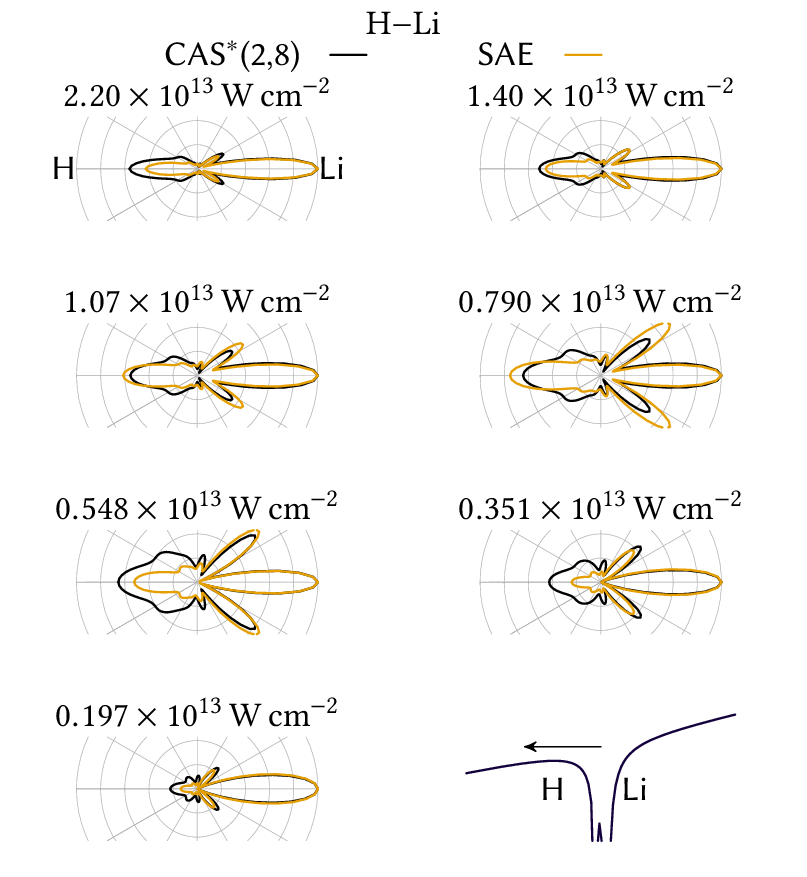}
 \caption{(color online). Molecular PADs at different field strengths showing the converged correlated simulations with a \cas{2}{8} and with the SAE approximation for the two 
 orientations of the molecule (left two columns \confLi, right columns \confH). The arrows indicate the predominating direction of electron emission.
 }
 \label{fig:pad-sae-cas}
\end{figure*}

To single out the influence of electron-electron correlations, the PADs from correlated \cas{2}{8} calculation
and those from a SAE calculation are shown in Fig.~\ref{fig:pad-sae-cas}. 
For the \confLi configuration (left two columns in Fig.~\ref{fig:pad-sae-cas}), SAE (bright curve) underestimates the size of the side maxima, especially at intermediate field strengths (central panels). At smaller field strengths ($\unit[0.351\times 10^{13}]{W\,cm^{-2}}$ and $\unit[0.197\times 10^{13}]{W\,cm^{-2}}$), the maximum pointing away from the \ce{Li} nucleus is either over- or underestimated, showing that no general pattern can be reasoned from correlation effects in PADs at different intensities.

For the \confH configuration and for all field intensities (right two columns in Fig.~\ref{fig:pad-sae-cas}), SAE drastically underestimates the side maxima at the site of the \ce{H} nucleus. On the other hand, at intermediate intensities, the other side maxima are overestimated by the SAE approximation. 
Thus, the favored direction of emission of electrons is decided by the electron-electron correlation for intermediate field strengths (see Fig.~\ref{fig:meta}). 

\section{Conclusions}
In this paper we presented a time-dependent approach to correlated electron dynamics following the excitation of diatomic molecules with strong electromagnetic fields.
The method is based on the TD-GAS-CI approach using a prolate-spheroidal representation of the single-particle orbitals within a partition-in-space concept to 
allow for good convergence of the truncated CI expansion. Thereby, parts of the multi-particle wave function close-by the nuclei are represented within a Hartree-Fock-like
orbital basis and the ejected part is represented in a grid-like FE-DVR basis set.

We illustrated the method by its application to the calculation of angle-resolved photoelectron spectra of the four-electron heteronuclear LiH molecule
with and without taking electron-electron correlation contributions into account.
To demonstrate the capabilities of the present approach, we then concentrated on the strong-field ionization of LiH using single-cycle pulses.
The ionization yield for the two opposite orientations of the molecule along the linearly polarized electric field
was calculated and an intensity-dependent shift of the preferred configuration
was observed: While for low intensities in the tunneling regime, ionization for \confH is larger, for high intensities well above the barrier, \confLi
shows higher yields. In between both regimes, a smooth transition is found. By turning on electronic correlations in the simulation,
we find that especially yields in the tunneling regime are affected whereas the high-intensity regime is well described using the SAE or CIS approximations.
Correlations shift the preferred configuration from \confLi to \confH for
low intensities and vice versa for high intensities. In a certain intermediate intensity regime even an interchange of the preferred configuration in comparison to uncorrelated
calculations is observed.
Additionally, angle-resolved photoionization distributions were presented and discussed,
and the correlation effects were singled out by comparison to SAE calculations.

Our results demonstrate the importance of electron-electron correlations in strong-field excitation scenarios of
diatomic molecules. We expect the TD-GAS-CI approach in combination with the prolate spheroidal basis set to be applicable to 
larger systems such as the CO molecule and to arbitrary polarization of the exciting pulse in the near future,
where experimental data is available~\cite{li_orientation_2011,wu_multiorbital_2012}.
Further, the application to two-color excitation scenarios, such as streaking and XUV-XUV pump-probe, and the exploration of correlation effects in molecular systems on ultrashort time scales,
e.g., post-collision interaction effects~\cite{schuette_2012,bauch_2012} or the time-delay in photoemission is within reach.

\begin{acknowledgments}
The authors thank C. Hinz for indispensable optimizations of the TD-GAS-CI code.
The authors gratefully acknowledge discussions with L. B. Madsen.
H.~R.~Larsson acknowledges financial support by the ``Studienstiftung des deutschen Volkes'' and the ``Fonds der Chemischen Industrie''.
This work was supported by the BMBF in the frame of the ``Verbundprojekt FSP 302'' and computing time at the HLRN via grants \texttt{shp00006} and \texttt{shp00013}. 
\end{acknowledgments}

\appendix
 
\section{Matrix elements}
\label{app:matrix_elements}

Because, to our knowledge, the formulas of all needed matrix elements have not been stated in one single publication or exhibit some misprints,
we briefly summarize the equations to simplify their implementation.

The volume element and the Laplacian are 
 \begin{align}
 \dd V =& a^3 (\xi^2-\eta^2)\dd \xi \dd \eta \dd \phi\label{eq:dV},\\
 \Delta =& \frac{1}{a^2 (\xi^2 -\eta^2)} \left[\partd{}\xi (\xi^2-1)\partd{}\xi + \partd{}\eta (1-\eta^2)\partd{}\eta\right. \nonumber\\
 &+ \left.\frac{\xi^2-\eta^2}{(\xi^2-1)(1-\eta^2)} \partdd{}\phi\right],
 \end{align}

 The action of $\hat T = -\Delta/2$ on $\Psi^m$ becomes 
 \begin{align}
  \hat T \Psi^m =& \left[ -\frac{1}{2 a^2(\xi^2-\eta^2)}\left(\partd{}\xi (\xi^2-1)\partd{}\xi\right.\right.  \label{eq:action-T}\\
  &+\left.\left. \partd{}\eta (1-\eta^2)\partd{}\eta - \frac{m^2}{\xi^2-1} -\frac{m^2}{1-\eta^2}\right) \right] \Psi^m.\nonumber
\end{align}

 The factor $(\xi^2-\eta^2)^{-1}$ [e.g. Eq.~\eqref{eq:action-T}] 
is canceled by the volume element, which avoids numerical problems due to the singularities at $\xi = 1$, $\eta=\pm 1$. 
Still, for odd $m$ values, the exact eigenfunctions show non-polynomial behavior at these points and contain factors 
of $(\xi^2-1)^{\frac{|m|}{2}}(1-\eta^2)^{\frac{|m|}{2}}$~\cite{ponomarev_1976,meixner_book}.
Non-polynomial functions 
are poorly represented by \DVR, in which Gauss quadrature
is used. Therefore, for odd $m$, we multiply the basis functions $\theta_n$ by $\sqrt{(\xi^2-1)/(\xi_n^2-1)}$ or $\sqrt{(1-\eta^2)/(1-\eta_n^2)}$ 
to avoid a non-polynomial integrand~\cite{tao_2009}:
\begin{align}
y_n(x) &=  \frac1{\sqrt{\omega_n}} \prod_{i\neq n}^N \frac{x-x_i}{x_n-x_i},\label{eq:dvr_chi}\\
 \theta^{m,\xi}_n(\xi) &= y_n(\xi) \times %
 \begin{cases}
  1, & m \text{ even}\\
  \sqrt{\frac{\xi^2-1}{\xi_n^2-1}}, & m \text{ odd},
 \end{cases}\label{eq:dvr_xi}\\
  \theta^{m,\eta}_n(\eta) &= y_n(\eta) \times %
 \begin{cases}
  1, & m \text{ even}\\
  \sqrt{\frac{1-\eta^2}{1-\eta_n^2}}, & m \text{ odd}.
 \end{cases}\label{eq:dvr_eta}
\end{align}
$\xi_n$ ($\eta_n$) is the DVR grid point of the corresponding basis function.
These definitions differ in case of a bridge function in the FEDVR basis~\cite{balzer_2010,rescigno_2000}.

\subsection{Potential energy and dipole operator}
The potential, \eq~\eqref{eq:pot_diat}, is independent of $\phi$. Hence, the potential matrix elements are evaluated using the \DVR properties:
\begin{equation}
\begin{aligned}
 \matrixe{f_{ia}^m}{V}{f_{jb}^{m'}} \equiv& V^m_{ia,jb} \\
 =& -\frac1a\delta_{mm'}\delta_{ij}\delta_{
ab} \\
&\times \frac{\left[ (Z_1+Z_2)\xi_i + (Z_2-Z_1)\eta_a\right]}{\xi_i^2 - \eta_a^2}.
\end{aligned}
\end{equation}
The matrix elements of the dipole operator in $z$-direction are 
\begin{align}
 \matrixe{f_{ia}^m}{z}{f_{jb}^{m'}} = \matrixe{f_{ia}^m}{a \xi \eta}{f_{jb}^{m'}} = a \delta_{mm'} \delta_{ij}\delta_{ab} \xi_i \eta_a.
\end{align}
The term stemming from the volume element, \eq~\eqref{eq:dV}, is canceled by the normalization factor of the basis functions, \eq~\eqref{eq:basis_func_dev}.

\subsection{Kinetic energy}
The matrix elements for the kinetic energy are evaluated using integration by parts and the \DVR quadrature \cite{tao_2009}:
\begin{align}
\begin{split}
 2a^2 \zeta T^{m,m'}_{ia,jb} =&\delta_{mm'}\delta_{ab} \int_{1}^\infty \dd \xi (\xi^2-1) \partd{\theta^m_i}\xi \partd{\theta^m_j}\xi\\ 
 &+ \delta_{mm'}\delta_{ij}\int_{-1}^1 \dd \eta (1-\eta^2) \partd{\theta^m_a}\eta \partd{\theta^m_b}\eta \\
 &+ \delta_{mm'}\delta_{ij}\delta_{ab}\left(\frac{m^2}{\xi_i^2-1} + \frac{m^2}{1-\eta_a^2}\right),
 \label{eq:tmat}
 \end{split}\\
\zeta \equiv&  \sqrt{(\xi_i^2-\eta_a^2)(\xi_j^2-\eta_b^2)}.
 \end{align}

\subsection{Interaction energy}
The Coulomb interaction of the electrons in Eq.~\eqref{eq:Hamiltonian} can be decomposed using the well-known Neumann series in prolate spheroidal 
coordinates~\cite{Neumann_1848,Neumann_series_Ruedenberg_1951}:
\begin{widetext}
\begin{align}
\frac{1}{|\vec{r}_1-\vec{r}_2|} =&\frac{1}{r_{12}} = \frac{4\pi}{a}\sum_{l=0}^\infty \sum_{M=-l}^l (-1)^{|M|} \frac{(l-|M|)!}{(l+|M|)!} {P}^{|M|}_l(\xi_<){Q}_l^{|M|}(\xi_>) 
\times Y_l^{m}(\arccos(\eta_1),\phi_1) Y_l^{m*}(\arccos(\eta_2),\phi_2),
\\
\begin{split}
=& \frac1a \sum_{l=0}^\infty\sum_{M=-l}^l (-1)^{|M|} (2l+1) \left(\frac{(l-|M|)!}{(l+|M|)!}\right)^2  {P}^{|M|}_l(\xi_<){Q}_l^{|M|}(\xi_>)  P_l^{|M|}(\eta_1){P}_l^{|M|}(\eta_2)\times\exp\bigl(\ii M(\phi_1-\phi_2)\bigr),
\end{split}
\end{align}
\end{widetext}
with $\xi_<\equiv \min(\xi_1,\xi_2)$ and $\xi_>\equiv \max(\xi_1,\xi_2)$.
$Y_l^{m}$ are the spherical harmonics and $P_l^m$ ($Q_l^m$) are the (irregular) associated Legendre functions~\cite{abramowitz_book,arfkenweber}.
For their computation for $\xi>1$, the code from \Ref~\cite{legendre_f_segura_1998} was used.

The $\eta$- and $\phi$-dependent parts are evaluated straight-forwardly using the properties of the basis functions. Because the Legendre functions exhibit a singularity for $\xi \rightarrow 1$, the $\xi$-dependent parts are not evaluated by \DVR quadrature but by solving the corresponding differential equation of the Green's function $ P_l^{|M|}(\xi_<){Q}_l^{|M|}(\xi_>)$ \cite{haxton_2011,guan_2011}.
The final expression for the integrals is then
\begin{widetext}
\begin{align}
 \matrixe{f_{i_1a_1}^{m_1}f_{i_2a_2}^{m_2}}{r_{12}^{-1}}{f_{j_1b_1}^{m'_1}f_{j_2b_2}^{m_2'}} %
  =& \delta_{a_1b_1}\delta_{a_2b_2}\delta_{i_1j_1}\delta_{i_2j_2}  \delta_{m_1-m_1',m_2'-m_2}  \Omega_{i_1i_2,a_1a_2}^{|m_1-m_1'|},\label{eq:w12_final}\\
  \Omega_{i_1i_2,a_1a_2}^{|M|}  
  =&a^{-1}\sum_{l=|M|}^{l_\text{max}}   P_l^{|M|}(\eta_{a_1}) P_l^{|M|}(\eta_{a_2})(2l+1)\nonumber\\
  &\times\Biggl[ (-1)^{|M|}   \left( \frac{(l-|M|)!}{(l+|M|)!} \right)^2 P_l^{|M|}(\xi_{i_1}) P_l^{|M|}(\xi_{i_2})   \frac{Q_l^{|M|}(\xi_N)}{P_l^{|M|}(\xi_N)}- \frac{(l-|M|)!}{(l+|M|)!}\frac{[T_{i_1i_2}^{lM}]^{-1}}{\sqrt{\omega_{i_1}\omega_{i_2}}}   \Biggr],\nonumber\\
   T_{ij}^{lM} =& -\delta_{ij} \left(\frac{M^2}{\xi_i^2-1} + l(l+1)\right) - \int \dd \xi (\xi^2-1) \partd{\theta_i^M}\xi \partd{\theta_j^M}\xi.\label{eq:w12_tmat}
  \end{align}
  \end{widetext}
$\xi_N$ is the value of the last (excluded) \FEDVR grid point, and $\omega_i$ are the quadrature weights of the \FEDVR grid points.
In the implementation, the Legendre polynomials in $\eta$ and the term in brackets are precomputed and stored in arrays so that the
actual computation of the matrix elements is just summing up the product of three array values. The matrix elements are symmetric 
in $\xi$ and $\eta$, which can be exploited as well. 

Since the integration over $\eta$ is still handled by usual quadrature, the maximum used value of $l$ in the series 
expansion, $l_\text{max}$, should not be too large such that the integration kernel for the $\eta$-dependent part is not a polynomial of
degree $2N_\eta-1$ any more ($N_\eta$ is the number of functions in $\eta$). Numerically, however, the results are not 
very sensitive to the choice of $l_\text{max}$ and setting it to the number of used basis functions in $\eta$ leads to good
results~\cite{haxton_2011,guan_2011}.

\section{Integral transformation}
\label{app:integral_trafo}
The usage of a DVR-based basis with its diagonality of potential and, to some extent, interaction matrix elements and the utilization of a partially
rotated basis allows for a massive reduction of the usual scaling relationships in quantum-chemical algorithms. In the following, 
the most crucial ones, namely the generation of the Fock matrix in Hartree-Fock and the integral transformation to the basis of molecular orbitals are shown. For convenience, we use the ``chemist's''
notation~\cite{szabo_ostlund_book} of the electron-electron repulsion integrals; \ie, $\smatrixe{ik}{r_{12}^{-1}}{jl} \equiv (ij|kl)$.
\subsection{Coulomb matrix in Hartree-Fock}
The bottleneck in usual Hartree-Fock calculations is the generation of the Coulomb and the exchange matrix as parts of the Fock matrix. 
The elements of the Coulomb matrix are generated by
\begin{align}
 J_{ij} &= \sum_{kl} D_{kl} (ij|kl).
\end{align}
$\matr D$ is the density matrix, \eq~\eqref{eq:density_matrix}. The scaling with the number of basis functions $N_b$ is $\mathcal O(N_b^4)$ if no additional techniques 
like density fitting are used. For the exchange matrix, the summation runs over the integral $(ik|lj)$ and the optimization procedure is similar. 

Resolving the indices using Eqs.~\eqref{eq:basis_func_dev} and \eqref{eq:w12_final}, 
\begin{align}
\begin{split}
 J_{ij} =& \sum_{m_ki_ka_k} \sum_{m_li_la_l}D_{kl} (\bfunc{i}\bfunc{j}|\bfunc{k}\bfunc{l}),\\
 =&  \sum_{m_ki_ka_k} \sum_{m_li_la_l}D_{kl} \delta_{a_ia_j}\delta_{a_ka_l}\delta_{i_ii_j}\delta_{i_ki_l} \\
 &\times \delta_{m_i-m_j,m_l-m_k} \Omega^{|m_i-m_j|}_{i_ii_k,a_i,a_k},\\
 =&\;\; \delta_{a_ia_j} \delta_{i_ii_j}   \sum_{m_k=m_k^s}^{m_k^e}\sum_{i_ka_k}  D_{kl}  \Omega^{|m_i-m_j|}_{i_ii_k,a_i,a_k},\label{eq:coulomb_sumcalc}\\
 \end{split}\\
 m_k^s =& \begin{cases}
           -m_\text{max} - (m_i - m_j),& m_i - m_j < 0\\
           m_\text{max}, & \text{else},
          \end{cases}\\
 m_k^e =& \begin{cases}
           -m_\text{max},& m_i - m_j < 0\\
          m_\text{max} - (m_i-m_j), & \text{else}.
          \end{cases}
\end{align}
The last two $\delta$-symbols can be resolved in the overall loop creating the Coulomb matrix. Therefore, an overall scaling of less than 
$N_b^2 N_\phi$ is achieved for the construction of $\matr J$, where $N_\phi$ is the number of basis functions in $\phi$.
The diagonalization of the Fock matrix is then the bottleneck of the SCF procedure. 

\subsection{Integral transformation to the partially rotated basis}
\label{subsec:integral_trafo_part_rot_bas}
The structure of the coefficient matrix for a partially rotated basis is~\cite{hochstuhl_2012,bauch_2014}
\begin{equation}
 \matr C = \begin{pmatrix}
      \mathbf{C_\text{rot}} & 0 \\
      0 & \mathbf{1} \\
     \end{pmatrix},\label{eq:mixed_bas_coeffmat}
\end{equation}
where $\mathbf{C_\text{rot}}$ is a dense matrix of size $N_\text{rot}\times N_\text{rot}$ for the rotated block of basis functions.

\subsubsection{Integrals between rotated orbitals}
\label{sec:hf_integrals}
In the following, orbitals indexed with $p,q,r$ or $s$ denote rotated orbitals (from the inner spatial region) and those indexed with $a,b,c$ or $d$ denote nonrotated
orbitals from the inner region that have to be rotated. 
The rotated orbitals are unitarily transformed by the real-valued coefficient matrix $\matr C$, \eq~\eqref{eq:mixed_bas_coeffmat}.
The integrals in the rotated molecular orbital frame are then~\cite{szabo_ostlund_book}
\begin{align}
 \smatrixe{p}{\hat h}{q} &= \sum_{ab}^{N_\text{rot}} C_{ap} C_{bq} \smatrixe{a}{\hat h}{b},\label{eq:2idx_trafo}\\
 (pq|rs) &= \sum_{abcd}^{N_\text{rot}} C_{ap} C_{bq} C_{cr} C_{ds} (ab|cd).\label{eq:4idx_trafo}
\end{align}
Because of the structure of the coefficient matrix, \eq~\eqref{eq:mixed_bas_coeffmat}, the sum runs only over all rotated 
orbitals, if $\{p,q,r,s\}$ are all themselves  rotated orbitals.
It is well known that the formally $\mathcal O(N_\text{rot}^8)$-scaling transformation 
can be massively reduced by employing partial transformations~\cite{szabo_ostlund_book,bender_1972,fink_1974,taylor_1987}, i.\,e.,
first transforming and storing the fourth orbital to $(ab|cs)$, then, transforming the third orbital to $(ab|rs)$ and so on. This scales as
$\mathcal O(N_\text{rot}^5)$ but twice the memory is needed for storing the intermediate transformed integrals. Doing only two two-index
transformation steps saves a considerable amount of memory but the scaling becomes worse. However, this is sometimes favorable~\cite{lmp2_werner_1998}.
A similar procedure can be applied for the one-electron integrals, \eq~\eqref{eq:2idx_trafo}.

In our case  however, the transformation can again be sped up massively, as exemplified by the calculation of the Coulomb matrix, see
\eq~\eqref{eq:coulomb_sumcalc}. Because of the diagonality in $\xi$ and $\phi$, it is beneficial to first transform the last two indices 
like in \eq~\eqref{eq:coulomb_sumcalc}, where $D_{kl}$ has to be replaced with $C_{ka}C_{lb}$. Note that, although the  integrals are real-valued, 
the orbitals for $m_\text{max}\neq 0$ are not:
\begin{align}
 (\bfunc{i}\bfunc{j}|\bfunc{k}\bfunc{l}) &\equiv (ij|kl) \neq (ij|lk) \\
 (ij|lk) &\equiv  (\bfunc{i}\bfunc{j}|\bfunc{l}\bfunc{k})\\
 (\bfunc{i}\bfunc{j}|\bfunc{k}\bfunc{l}) &= (\bfunc{i}\bfunc{j} |f_{i_la_l}^{-m_l}f_{i_ka_k}^{-m_k} ).
 \end{align}
 Hence, only the following symmetries hold:
 \begin{align}
    (ij|kl) = (kl|ij) = (ji|lk) = (lk|ji) \neq (lk|ij) \neq \dots \label{eq:twoelints_symmetry_complex}
\end{align}
Because the two-index transformed integrals have even less symmetry, $N_\text{rot}^3\times (2m_\text{max}+1)$ elements need to be stored for them.
The transformation of the remaining two indices are done using partial transformations, as usual in quantum chemistry (see above). For large non-rotated bases, the fully transformed 
integral tensor requires too much memory and the latter transformation is done on the fly: 
\begin{align}
 (pq|rs) &= \sum_{a=1}^{N_\text{rot}} \sum_{m_b= -m_\text{max}}^{m_\text{max}} C_{ap} C_{bq} (ab|rs).
\end{align}
The index $b$ on the right-hand-side is constructed by $m_b$ and the indices for the functions in $\xi$ and $\eta$ of the index $a$.
The product $C_{ap}C_{bq}$ can be precalculated which simplifies and accelerates the summation. Transforming the last two indices on the fly 
requires $N_\text{rot}\times (2m_\text{max}+1)$ operations, which is still better than the usual requirement of $N_\text{rot}^2$ operations for a non-DVR basis such that 
the overall scaling for the complete two index transformation is then $N_\text{rot}^5\times(2m_\text{max}+1)$ instead of the best achievable $N_\text{rot}^4$ scaling.

\subsubsection{Integrals between nonrotated orbitals and rotated orbitals}
\label{subsec:integrals_nonrot_rot}
If all basis functions are nonrotated \DVR functions, the formulas for the one- and two-electron integrals, \eq~\eqref{eq:w12_final},
 can be applied directly. The crucial point is the efficient implementation of the integrals between both 
nonrotated and rotated basis functions. A lot of simplifications come from the structure of the coefficient matrix, \eq~\eqref{eq:mixed_bas_coeffmat}, 
which is diagonal if both basis functions are nonrotated functions and zero if one function is a rotated and another a nonrotated function. 
In the following, nonrotated basis functions are underlined. 

The mixed one-electron integrals are then:
\begin{align}
 \smatrixe{p}{\hat h}{\underline q} &=\smatrixe{\underline q}{\hat h}{p} =\sum_{ab}^{N_b} C_{ap} C_{b\underline q} \smatrixe{a}{\hat h}{b},\\
 C_{b\underline q} &= \delta_{b\underline q},\\
\Rightarrow \smatrixe{p}{\hat h}{\underline q} &= \sum_{a}^{N_\text{rot}} C_{ap} \smatrixe{a}{\hat h}{\underline q}.
 \end{align}
Since the potential is diagonal, $\smatrixe{p}{\hat V}{\underline q}=0$. The structure of the kinetic energy matrix [diagonality for $\phi$ functions,
see \eq~\eqref{eq:tmat}, and a banded sparsity pattern] can be exploited as well.

For the two-electron integrals, several cases have to be considered. 
\paragraph{One nonrotated function:} If $s > N_\text{rot}$, \eq~\eqref{eq:4idx_trafo} reduces to
\begin{align}
 (pq|r\underline s) &=  \sum_{abcd}^{N_b} C_{ap} C_{bq} C_{cr} C_{d\underline s} (ab|cd),\\
 C_{d\underline s} &= \delta_{d\underline s}, \\
\Rightarrow (pq|r\underline s) &= \sum_{abc}^{N_\text{rot}} C_{ap} C_{bq} C_{cr} (ab|c\underline s).
\end{align}
Because $c \leq N_\text{rot}$, but $\underline s >N_\text{rot}$, all integrals $(ab|c\underline s)$ are zero. This originates from the 
diagonality of the $\xi$-functions. For symmetry reasons, this applies as well to $(pq|\underline r s) = (p\underline q|r s)$ and so on, see \eq~\eqref{eq:twoelints_symmetry_complex}.

\paragraph{Two nonrotated functions:} If $p$ and $r$ are nonrotated functions, the integrals are also zero:
\begin{align}
 (\underline p q|\underline r s) &= \sum_{bc}^{N_\text{rot}} C_{bq} C_{cr} \underbrace{(\underline p b| \underline r c)}_{=0} = 0 .
\end{align}
This changes if $r$ and $s$ are nonrotated functions:
\begin{align}
 (p q|\underline r \underline s) &= \sum_{ab}^{N_\text{rot}} C_{ap} C_{bq} (a  b| \underline r \underline s).
\end{align}
This sum is computed very efficiently, see section \ref{sec:hf_integrals}. Thus, we do not store these integrals but compute them on the fly. 
\paragraph{Three nonrotated functions:} The integrals are zero:
\begin{align}
 (p\underline q| \underline{rs}) = \sum_{a}^{N_\text{rot}} C_{ap} (a\underline q|\underline{rs}) = 0
\end{align}

To summarize, only if two nonrotated basis functions are used for the same electron, the mixed integrals are nonzero but can be computed in a 
very efficient manner exploiting the diagonality inherent to the underlying \DVR basis in $\xi$ and $\eta$. Therefore, the number of nonrotated basis 
functions for the GAS-CI-code do not influence the computational costs of the integrals considerably (but the number of configurations in the CI expansion).  
The computation of the two-electron integrals in  the ``raw'' \DVR basis is negligible.

\end{document}